# FDTD-based optical simulations methodology for CMOS image sensor pixels architecture and process optimization

Flavien Hirigoyen *[a], Axel Crocherie [a,b], Jérôme Vaillant [a] and Yvon Cazaux [a]
[a] STMicroelectronics, 850 rue Jean Monnet, 38920 Crolles, France
[b] Photonics, Electromagnetism, and Microelectronics Institute (IMEP), Grenoble (France)
*flavien.hirigoyen@st.com; phone: (33) 4 76 92 54 64; fax: (33) 4 76 92 56 78; www.st.com

## ABSTRACT

This paper presents a new FDTD-based optical simulation model dedicated to describe the optical performances of CMOS image sensors taking into account diffraction effects.

Following market trend and industrialization constraints, CMOS image sensors must be easily embedded into even smaller packages, which are now equipped with auto-focus and short-term coming zoom system. Due to miniaturization, the ray-tracing models used to evaluate pixels optical performances are not accurate anymore to describe the light propagation inside the sensor, because of diffraction effects. Thus we adopt a more fundamental description to take into account these diffraction effects: we chose to use Maxwell-Boltzmann based modeling to compute the propagation of light, and to use a software with an FDTD-based (Finite Difference Time Domain) engine to solve this propagation.

We present in this article the complete methodology of this modeling: on one hand incoherent plane waves are propagated to approximate a product-use diffuse-like source, on the other hand we use periodic conditions to limit the size of the simulated model and both memory and computation time. After having presented the correlation of the model with measurements we will illustrate its use in the case of the optimization of a 1.75µm pixel.

## 1. INTRODUCTION

The image sensors market has experienced considerable growth over recent years due to the increasing demands of digital still and video cameras, security cameras, webcams, and mainly mobile cameras. But the companies addressing this market must satisfy a strong price pressure while providing improved image quality and higher resolution sensors. Moreover these sensors must be easily embedded into smaller and smaller packages, which are now equipped with auto-focus and short-term coming zoom system [1, 2, 3 and 4]. Charge-coupled devices have been the dominant image sensor technology; nevertheless Complementary Metal Oxide Semiconductor (CMOS) technology has reached competitive performance and show advantages in particular regarding on-chip functionality and power consumption [5, 6 and 7]. A CMOS image sensor is composed of an array of pixels (light sensitive elements) surrounded by a read-out circuitry. Each pixel is made of a photodiode which collects the photogenerated electrons and of transistors dedicated to collect, convert and read these electrons. Organic color filters are placed above the pixels for color reconstruction [8], and microlenses are placed on top to concentrate light flux to the photodiode (see Figure 1 below).

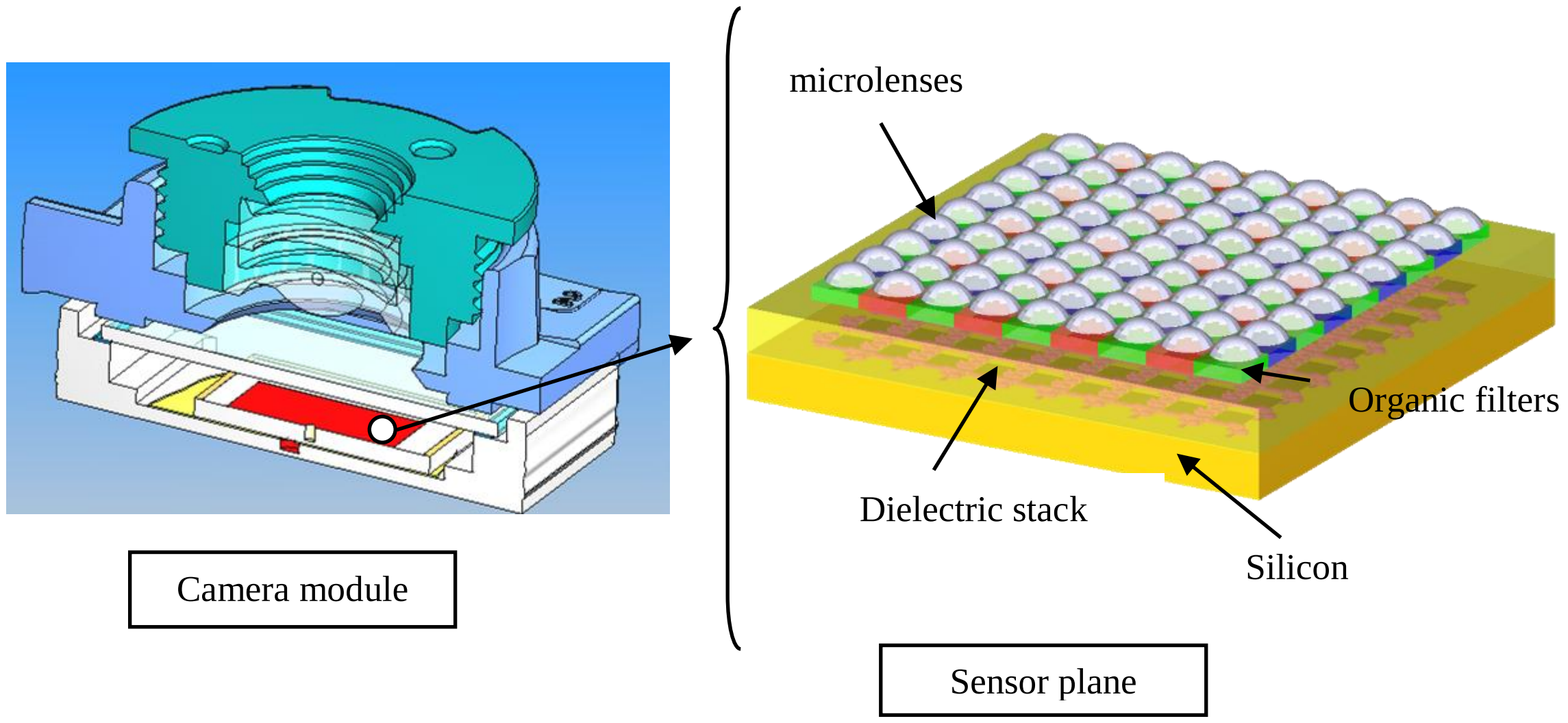

Figure 1: CMOS sensor in a camera module

To miniaturize the sensors the size of the pixels is reduced leading designers to share transistors between several pixels, to keep an acceptable fill factor (ratio of the photodiode area over the total pixel area). Thus microelectronics designers and process engineers need optical information to optimize in parallel both pixels architectures and processes, to maximize photon collection and to reach the performances required by the customers.

We previously developed at STMicroelectronics a ray-tracing based methodology to describe the optical efficiency of different architectures and processes. Using Snell-Descartes and Beer-Lambert description, we could simulate the exact layout and process of the pixels under product-like illumination. This methodology was fast and easy to use thanks to the ray-tracing algorithm itself [10]. Unfortunately, the ray-tracing physics is not accurate anymore to describe the light propagation inside the pixels, because the diffraction-limit size in the visible (3µm) is now reached with the miniaturization.

Among several alternatives we chose to adopt a more fundamental description to take into account these diffraction effects: we use Maxwell-Boltzmann based modeling to compute the propagation of light, and an FDTD-based (Finite Difference Time Domain) engine to solve this propagation.

We present in this article the complete methodology. Firstly we will show the limitations of our precedent ray-tracing model, then we will introduce the FDTD principle based on Yee algorithm (see section 2.2). Secondly we will show how we customized the chosen software: on one hand from a source point of view we approximate a product-use diffuse-like source, on the other hand we use periodic conditions to limit the size of the simulated model and both memory and computation time (see section 2.4). In a last part we will present the validation of this methodology from the comparison of the simulated and measured angular responses of a pixel to a collimated beam. We will also illustrate the use of this simulation model with the optimization of a 1.75µm pixel [17]. Finally we will anticipate on future aperture effect on photon collection inside next sensors.

## 2. METHODOLOGY

### 2.1 From ray tracing to electromagnetism

CMOS image sensors pixels are designed using standard microelectronics CAD software. Unfortunately, these tools do not contain any optical module, thus we developed internally at STMicroelectronics our own optical simulations methodologies [10]. We adopted in 2002 a ray-tracing modeling based on the commercial software Zemax®[9] traditionally used to design optical systems such as telescopes, lenslets for zooms, etc… We created a tool to import the exact layout of the pixels from microelectronics CAD software and to make it compatible with the three-dimensional description of the optical simulator.

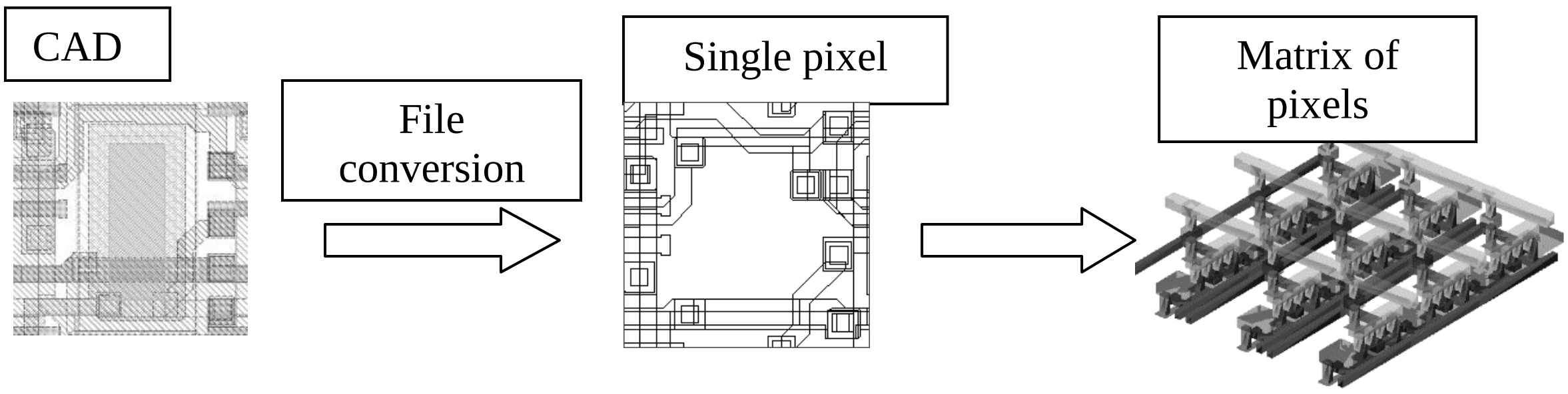

Figure 2: Model generation of pixel from CAD description to three-dimensional optical model

The optical parameters (refractive index and extinction coefficients) of the silicon and the dielectric layers containing these pixels were experimentally measured and inserted in the simulator, allowing us to propagate the light according to Snell-Descartes description (geometric aspect, equation 1) and Beer-Lambert description (energetic aspect, equation 2) :

$$n_i \sin(\theta_i) = n_r \sin(\theta_r) \qquad (1)$$

with $n_i$, $\theta_i$, $n_r$, $\theta_r$, refraction index and angle in incident and refractive layer respectively.

$$IT = e^{-\alpha\tau} \qquad (2)$$

with $IT$, internal transmittance, $\tau$ corresponding thickness, $\alpha$ absorption coefficient.

We validated this ray-tracing methodology by comparing the simulated and measured angular response of a pixel. It consists in illuminating the sensor plane by a collimated beam. At silicon level inside the pixels, the optimized microlenses focus the light into a sub-micron spot. Thus by rotating the sensor, we obtain their horizontal and vertical angular responses (see Figure 3).

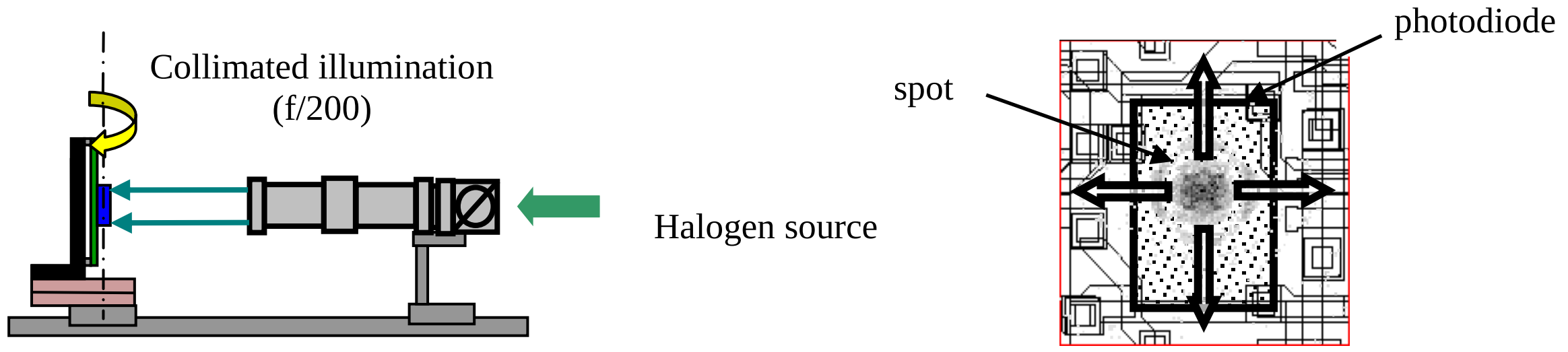

Figure 3: Angular response

Figure 4 shows an example of a simulated and measured angular response for two different pixels: a 5.6µm pixel and a more recent 3µm pixel.

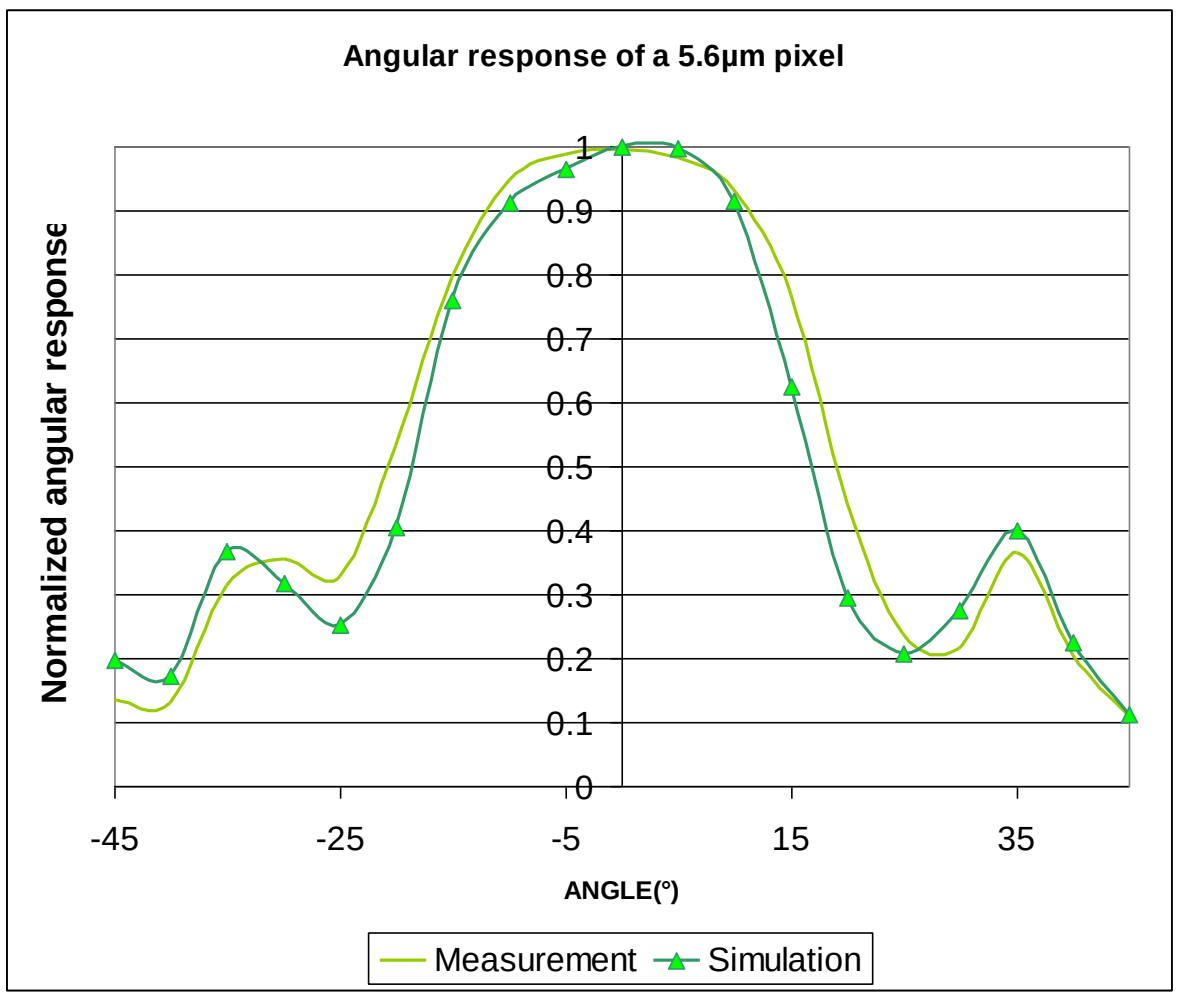

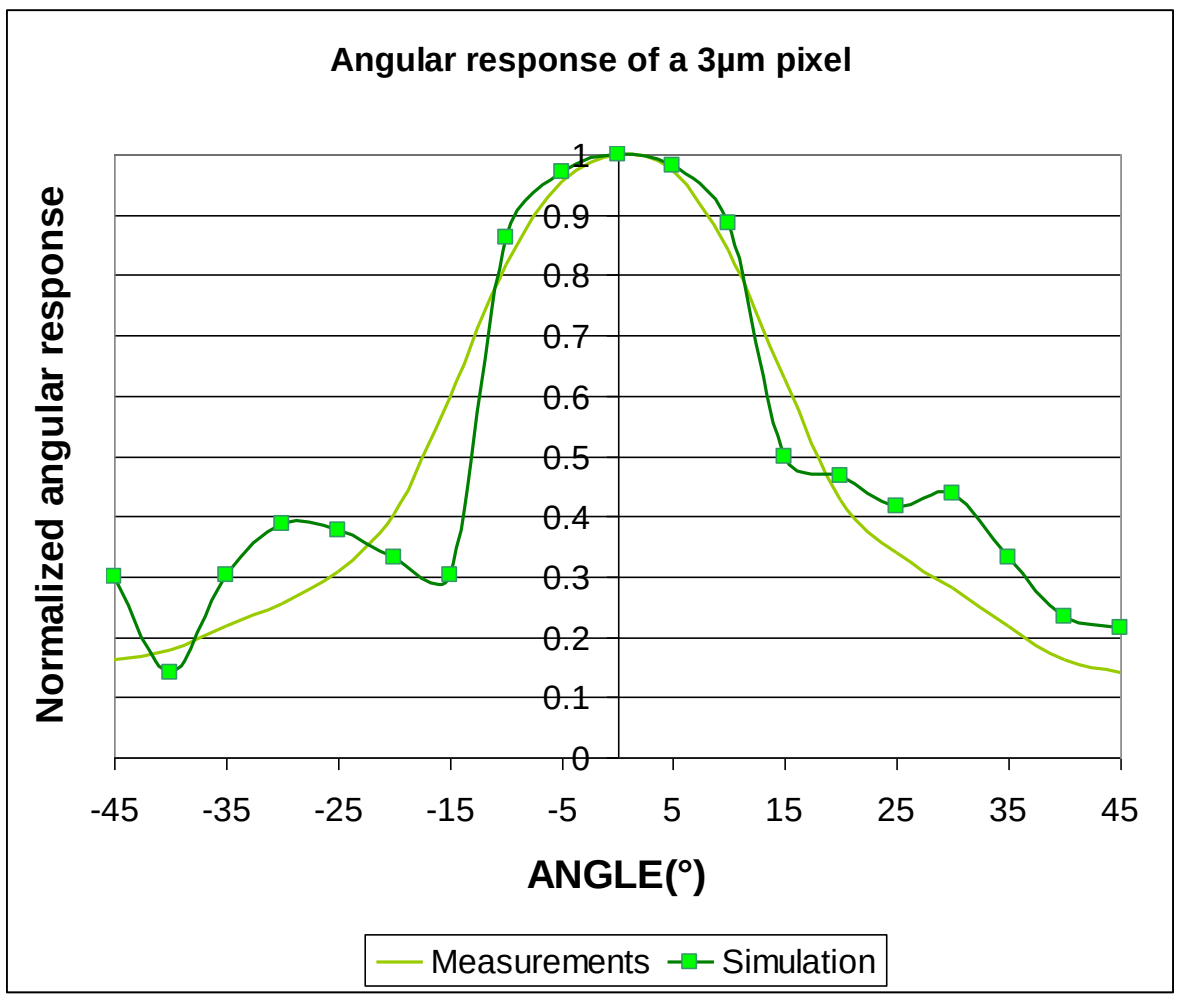

Figure 4 : Ray tracing angular responses vs. measurements: scaling effects

We can see a very good agreement between simulated and measured angular responses for the 5.6µm pixel, emphasizing the accuracy of the ray-tracing model. But for the 3µm pixel pitch, the same kind of simulation is not as well correlated. We have steep oscillations of the simulation response corresponding to the sharp shading effects of the architecture, whereas the measured response is smoother. This is due to diffraction effects which can not be neglected anymore at this scale: they tend to enlarge the focused spot at silicon level, and modify the energetic distribution compared to a ray-tracing modeling. This means that this model is not accurate to describe light propagation for pixels sizes inferior to 3µm: for a fixed architecture a Snell-Descartes description will propagate the light the same way whatever size, i.e. for a 30µm pixel as for a 3µm pixel. To overcome this we need a more fundamental description to take into account the diffraction effects. Among several possibilities we chose to use an electromagnetic simulation tool based on Finite Difference Time Domain.

## 2.2 FDTD principle

Finite Difference Time Domain methods consist in solving temporally the time-dependant Maxwell-Boltzmann equations in their partial differential form [11]:

$$\frac{\partial E}{\partial t} = \frac{1}{\varepsilon}(\nabla \times H - J) \tag{3}$$

$$\frac{\partial H}{\partial t} = -\frac{1}{\mu}(\nabla \times E) \tag{4}$$

With $E$, electric field, $H$ magnetic field, $\varepsilon$ permittivity, $\mu$ permeability of the medium and $J= \sigma E$ current density, σ conductivity of the medium. We can see that any change in the E-field in time is linked to any change in the H-field across space and vice versa.

Generally FDTD engines use Yee algorithm to solve these equations [12 and 13]: the simulated structure is sampled as Yee cells which allow a spatial and temporal discrete solve of these equations (see Figure 5).

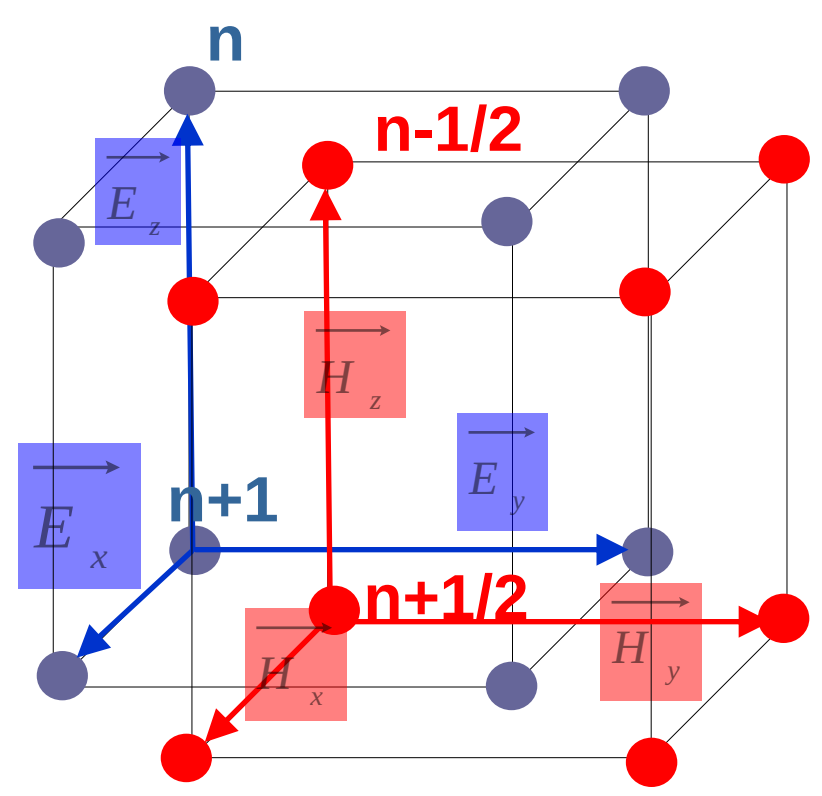

Figure 5: Yee cell scheme

These Yee cells consist in two interlaced grids: the electric components of the field are calculated at primary grid nodes, the magnetic components at the secondary grid nodes:

$$E^{n+1} = E^{n} + \frac{\Delta t}{\varepsilon}\left(\nabla \times H^{n+1/2} - \sigma E^{n}\right) \tag{5}$$

$$H^{n+1/2} = H^{n-1/2} - \frac{\Delta t}{\mu}\left(\nabla \times E^{n}\right) \tag{6}$$

Thus we obtain a system in which each new value of a field vector component at any Yee lattice point is expressed by its previous value and by the components of the other field vector at adjacent points. The process is repeated until an electromagnetic steady state solution is fully accomplished. The computation time step $\Delta t$ must be chosen so that the field is not expected to change significantly between two nodes, which mean in equivalent that the distance between two successive nodes must be inferior to a fraction of the propagating wavelength. This condition implies very large structures in terms of memory, particularly for 3D modeling.

To solve mathematically the system (5) and (6), we need boundary solutions, or boundary conditions. Two kinds of conditions exist: absorbing boundary conditions, periodic boundary conditions. The first one requires a sufficient large structure to get a relevant model while the other one considers the simulated structure as infinitely repeated, like a periodic crystal, which allows to reduce the simulated structure to its elementary zone. In that case the simulator solves the equations according to Bloch's Theorem.

The point of this work was to adapt these FDTD methodologies to get a relevant pixel model while keeping reasonable time computation and memories. Among several FDTD tools, we chose to adopt an optics dedicated tool *Lumerical FDTD solutions®* [14], because of the efficiency of its FTDT calculation engine, the multiple possibilities regarding optics design and the friendly ergonomics of the CAD interface.

### 2.3 Pixel model

A CMOS sensor pixel cumulates the functions of photon collection, photon-electron conversion, and finally reading. So a pixel is constituted by several complex elements, from a geometrical point of view and from a material point of view. Considering their optical influence we will separate these elements in three groups. Firstly we consider the planar layers: the silicon substrate, then the dielectric layers necessary to isolate the metallic interconnections elements. Secondly we have the pixel architecture positioned inside of them. At substrate surface we have the MOS transistors made of polysilicon bricks, then the metallic contacts and the alternate metal lines to drive the signal until the surrounding circuitry. Finally, we have the optical elements: the color filters array or Bayer pattern [8] constituted of colored resin dedicated to decompose the white light into its three red, green and blue primary colors, and above them the organic microlenses dedicated to focus light into the photodiode (see Figure 6).

If the design of the dielectrics layers is very simple under the CAD part of the software, the integration of the architecture can be very tedious. Thus a specific tool has been developed by Lumerical to import GDSII elements directly from microelectronics CAD software, so that it is not necessary to redesign each element.

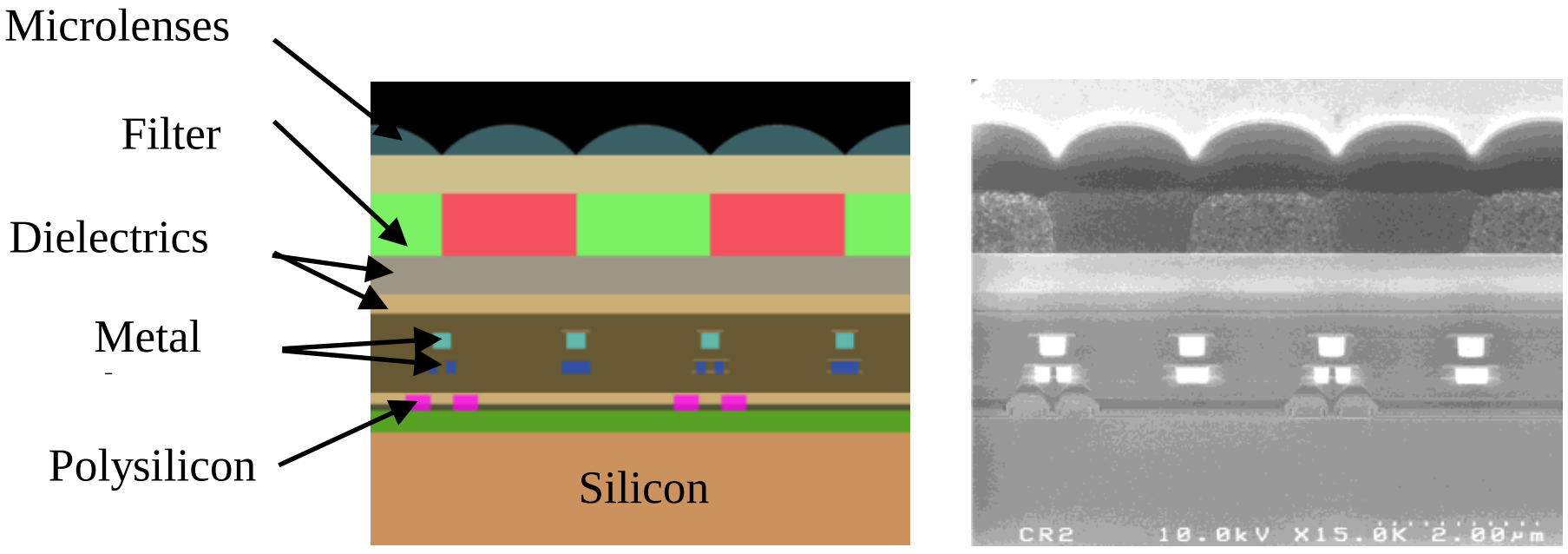

Figure 6: Structure of simulated pixels

Once geometrically defined, we must set the optical properties of each material. Since Maxwell-Boltzmann equations need permittivity and $\sigma$ conductivity of the materials to get solved we can define them by their refractive index *n* and extinction coefficient *k* thanks to the equivalence:

$$\varepsilon = n^2 \tag{7}$$

$$\sigma = 2nk\,\omega\varepsilon_0 \tag{8}$$

With $\omega$ angular frequency of the wave, $\varepsilon_0$ vacuum permittivity.

For the metal line case, we approximate them by perfect conductors: the simulator considers that their refraction index *n* is the same as the incident material where they are situated, and the conductivity $\sigma$ is infinite. In this way, any wave reaching the metal line is reflected with the same incident angle, and without any attenuation.

For the other materials we use a simple experimental data base of refractive index and extinction coefficients for pure monochromatic simulations, and dispersive models for multi-wavelengths simulations. Indeed FDTD formalism offers the possibility to extract multi-wavelengths results from a unique run by calculating the Fourier transforms of the propagated waves. In that case, the material parameters must be fitted on a dispersive model [15]. Thus each material permittivity must be decomposed in three available dispersive models: Debye, Lorentz and Plasma models defined in (9), (10), and (11) respectively.

$$\tilde{\varepsilon}_D(f) = \frac{\varepsilon_{DEBYE}\,.\upsilon_C}{\upsilon_c - i2\pi.f} \tag{9}$$

$$\tilde{\varepsilon}_L(f) = \frac{\varepsilon_{LORENTZ}\,.\varpi_0^2}{\varpi_0^2 - 2i\delta_0 2\pi.f - (2\pi.f)^2} \tag{10}$$

$$\tilde{\varepsilon}_P(f) = \frac{\varpi_p^2}{2\pi.f.(i\upsilon_c + 2\pi.f)} \tag{11}$$

where the coefficients $\varepsilon_{DEBYE}$, $\upsilon_c$, $\varepsilon_{LORENTZ}$, $\omega_0$, $\delta_0$, and $\omega_p$ are the fitting parameters to be determined. The total, complex-valued permittivity of the materials is given in (12).

$$\tilde{\varepsilon}(f) = \varepsilon_{real} + i.\varepsilon_{imag}\,.\frac{f_{sim}}{f} + i.\frac{\sigma}{2\pi.f\varepsilon_0} + \tilde{\varepsilon}_D(f) + \tilde{\varepsilon}_L(f) + \tilde{\varepsilon}_P(f) \tag{12}$$

The first two terms represent the contribution due to the background permittivity $\varepsilon$, the third term represents the contribution due to the conductivity $\sigma$, and the fourth through sixth terms represent the Debye, Lorentz and Plasma contributions discussed previously. Besides, $f_{sim}$ is the center frequency of the simulation.

### 2.4 Source

The source development is a key part of this methodology. Indeed, the objective of this optical simulation methodology is to correctly evaluate the pixels optical performances. This induces a source that reproduces a product-like illumination, keeping reasonable computation time and memory. This problem was already addressed while developing the ray-tracing based optical methodology [10]. Regarding scaling problems between the objective-lens (several millimeters) and the pixels (several micrometers), we chose to adopt a local approach and to simulate a group of pixels looking to the uniform exit pupil of the objective-lens. Thus their relative position inside the sensor was equivalent to the chief ray angle of this source. The ray-tracing source was composed of a uniform distribution of rays, spatially over the pixel area and angularly inside a cone defined by the f-number of the objective.

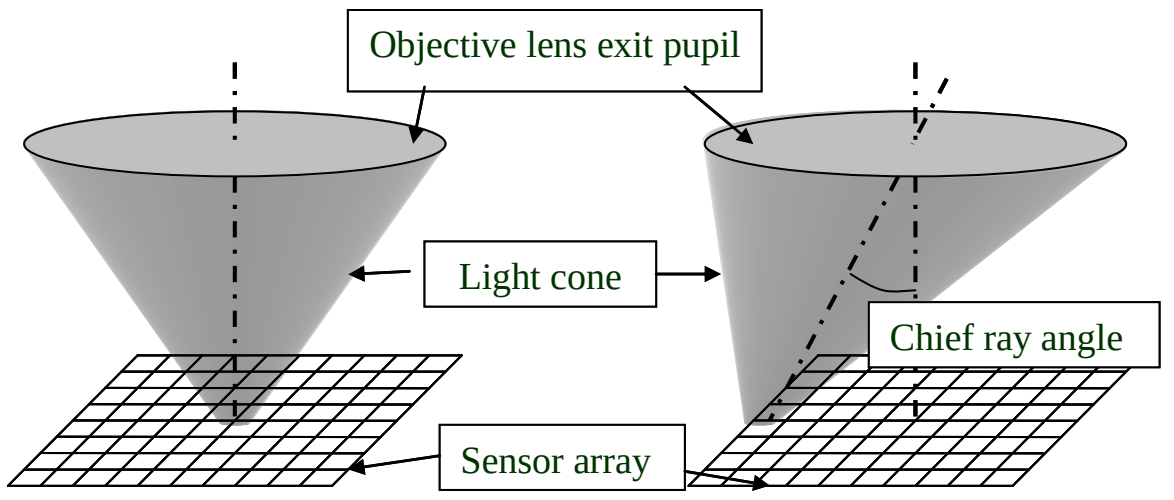

Figure 7: Diffuse-like source

We applied this approach to the electromagnetic modeling [16]. Since it is far away from the image plane, the object plane corresponds to a wide source which can be decomposed in a sum of different incoherent point sources emitting spherical waves. At entrance pupil we can approximate them by plane waves.

Therefore any image on the CMOS detector array can be reconstructed by incoherently summing the electromagnetic fields from these single point sources that pass through the objective-lens. Figure 8 below shows a schematic representation of the light source seen by the pixel array.

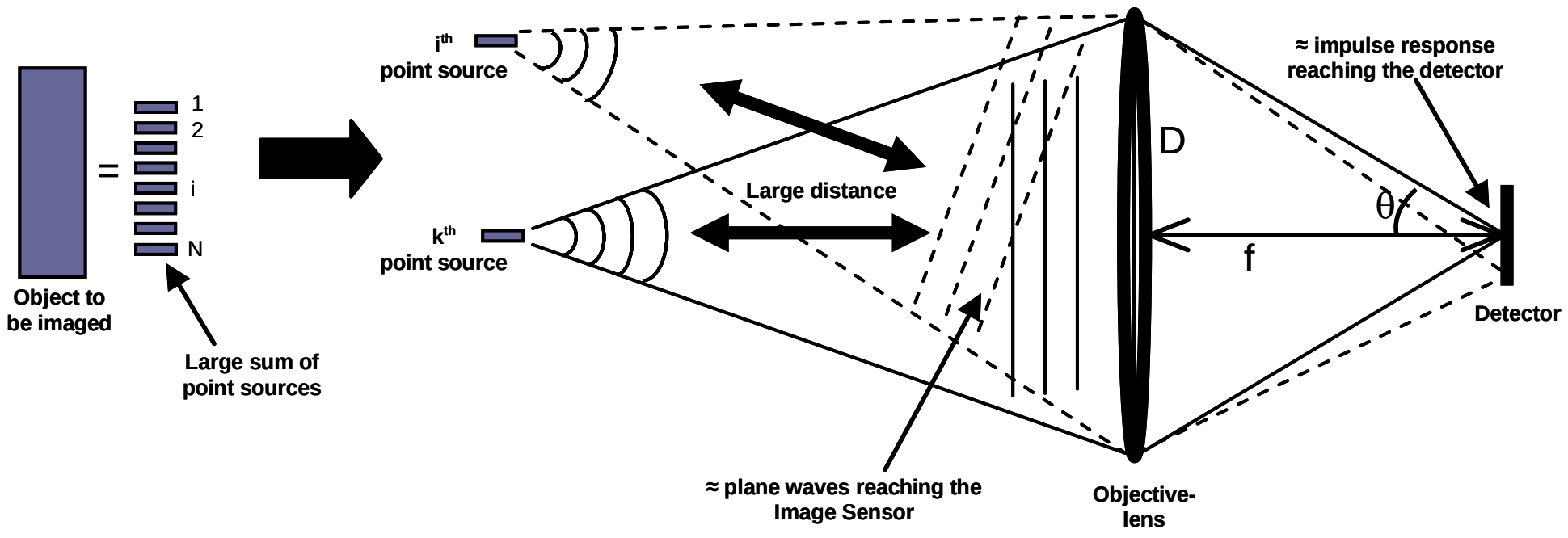

Figure 8: Wide source electromagnetic decomposition

A first source meets this approach in *Lumerical®*: it approximates the impulse response of an objective-lens by recreating a Gaussian or sinc source. To correctly simulate our uniform light source, we have to incoherently sum different spatial "thin lens" sources, translated along the exit pupil. But as we introduced it in section 2.2, FDTD simulations require boundary conditions, absorbing or periodic. Unfortunately since Gaussian or sinc sources are decomposed as a sum of different plane waves at different angles, periodic boundary conditions are unavailable, and we require absorbing boundary conditions leading to a wide structure. Thus it is not applicable for a three dimensional modeling because such a structure need a too large memory.

To overcome this, we developed another source modeling equivalent to the precedent. To simulate the incoherent illumination of the detector, we use plane waves instead of focused beams. A plane wave propagating at given incidence angle in a periodic structure can be implemented in FDTD using Bloch periodic boundary conditions, also known as the Sine-Cosine Method [11]. We have demonstrated that the incoherent sum of spatially uniformly distributed beams characterized by a given f-number is equivalent to the incoherent sum of angularly uniformly distributed plane waves with incidence angles limited by the previous f-number [16]. In this way it is possible to reduce the size of the simulated domain to the central one (one single pixel, or a complete Bayer pattern of 4 pixels).

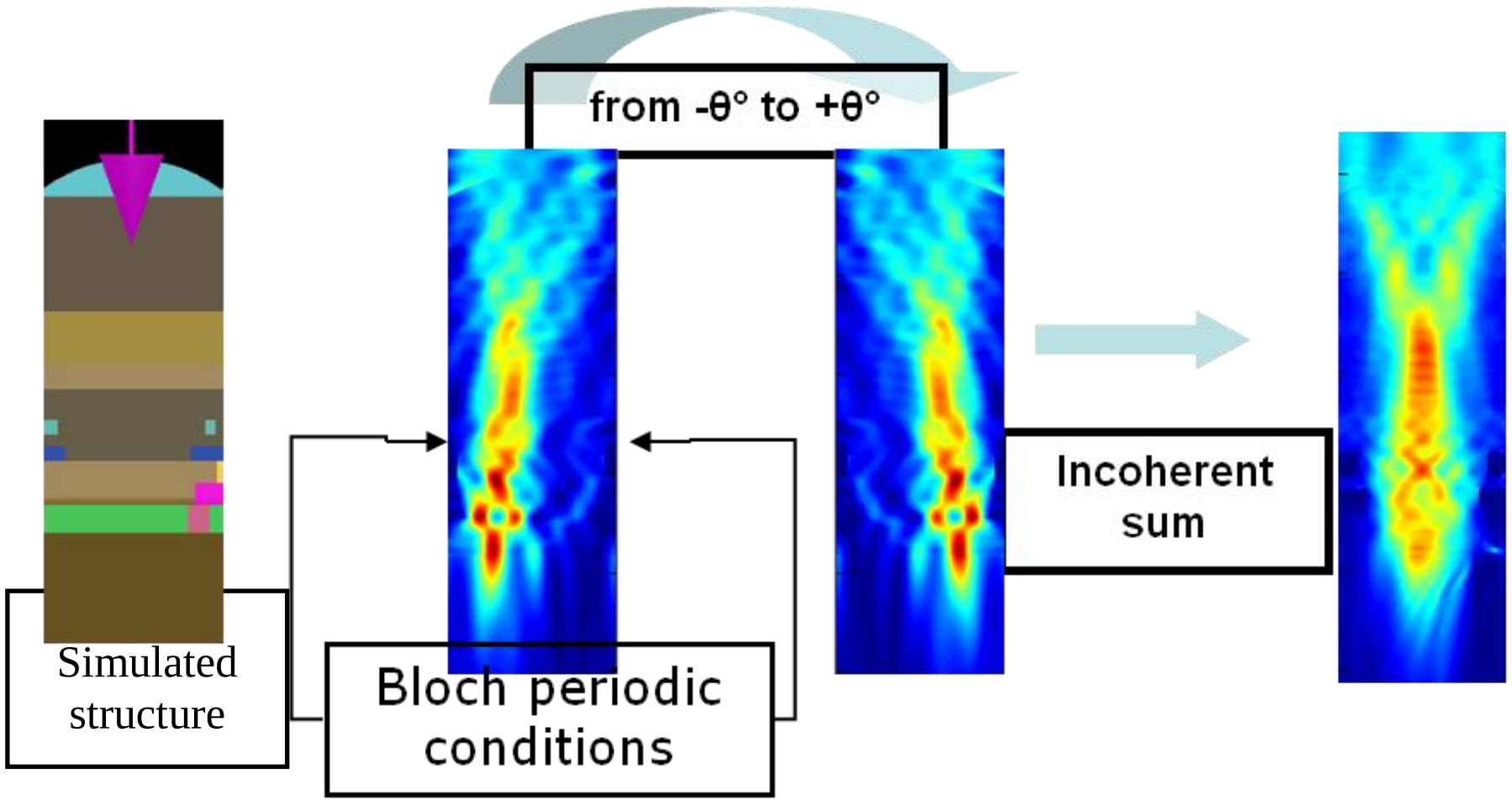

Figure 9: incoherent plane waves summation

Finally to obtain a complete diffuse-like illumination, we average s-polarized and p-polarized plane waves.

## 2.5 Optical generation

We have previously seen how to approximate a diffuse-like source by incoherently summing plane waves from different polarization states and angles. Now the information we want to extract is the energy distribution inside the pixel. It can be calculated from Poynting vector, which corresponds to the cross product of the electric vector and the magnetic vector:

$$\vec{P} = \frac{1}{2}.\left[\vec{E} \times \vec{H}\right] \tag{13}$$

with $P$, Poynting vector in W.m$^{-2}$, $E$, electric vector in V.m$^{-1}$, $H$, magnetic vector in A.m$^{-1}$.

The simulator propagates the electric field in the primary grid, and the magnetic field in the secondary grid. Then, once the steady state solution is accomplished, it interpolates the components of the magnetic field on the primary grid, so that it is possible to calculate the Poynting components anywhere inside the simulated region.

This Poynting vector is relevant if we want to compare the optical performance of a structure at back-end level, i.e. above the silicon level inside the dielectric stack. But if we want to quantify the overall optical performance, it is necessary to calculate the absorbed energy in the silicon, and to quantify the potential photogenerated electrons, or optical generation rate $G_{opt}$ (in m$^{-3}$.s$^{-1}$). This is done by calculating the divergence of the Poynting vector at silicon level:

$$G_{opt} = \eta . \frac{-\operatorname{Re}\left[\nabla \vec{P}\right]}{E_{ph}} \tag{14}$$

where $E_{ph}$ is the photon energy (in J) given by (15) and η is the quantum yield, which gives the average number of charge carriers generated by a single photon. In silicon at visible wavelengths, η=1 is assumed.

$$E_{ph} = h.\frac{c_0}{\lambda_0} \tag{15}$$

## 3. RESULTS

### 3.1 Methodology validation: angular response

As we have previously seen in section 2.1, the angular response is a good calibration methodology of the optical simulation model. The simulation consists in propagating plane waves at different angles, and to sum the optical generation in the silicon in the photodiode region. Then the results are normalized to the response at 0°. On Figure 10-a are plotted the measured and simulated angular responses of a 1.75µm pixel. We can see that the ray tracing simulation gives a steep response to a collimated beam, while the electromagnetic simulation is very similar to the measurement. This smoother response is induced by the diffraction phenomenon: the spot is expanded, and its energetic distribution is differently concentrated as it is illustrated on Figure 10-b , showing the spot at silicon level for both ray tracing and FDTD simulations.

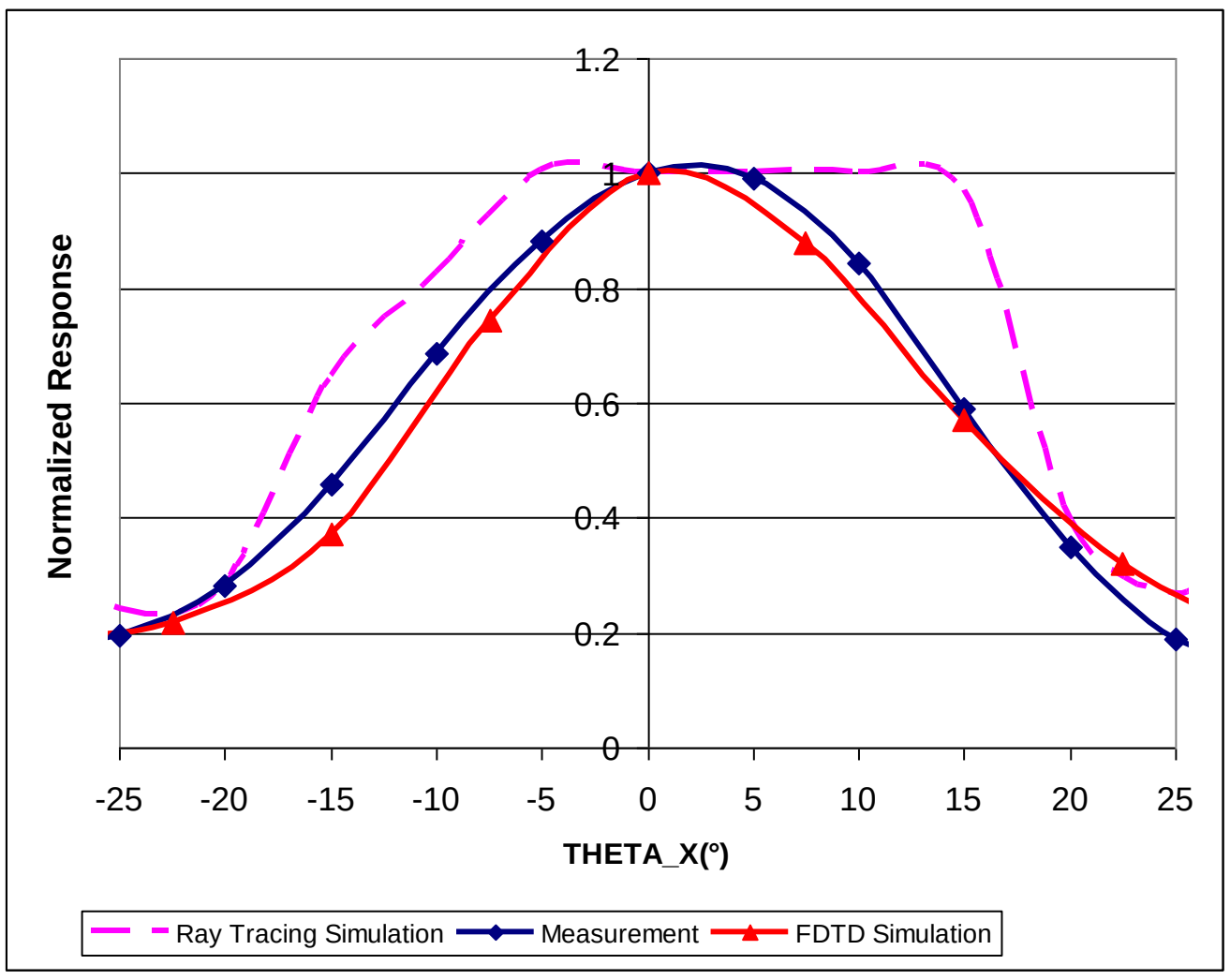

a

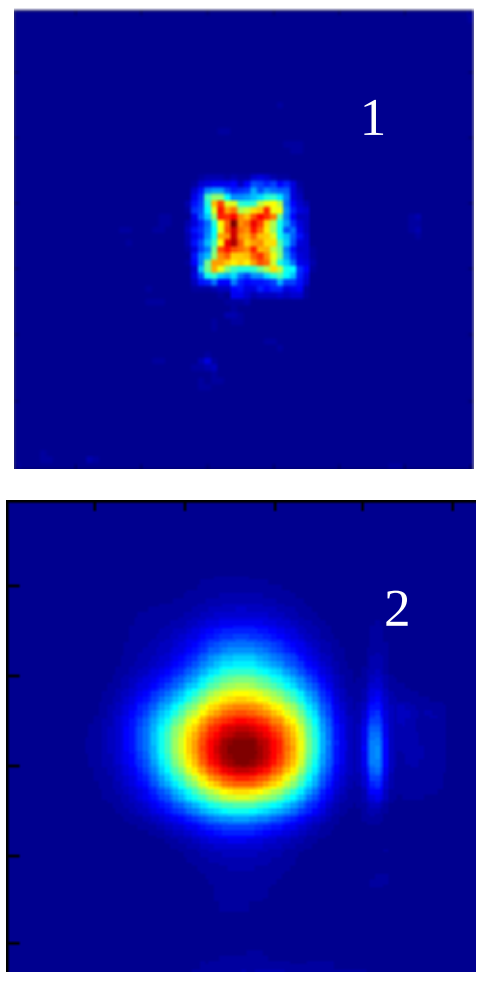

b

Figure 10: (a) Angular response of a 1.75µm pixel and (b) Spots at 0° at Si level, (1) ray tracing and (2) FDTD

### 3.2 Sensitivity improvement

Once this model calibrated, it is possible to predict optical performances of different architectures. Among other parameters, a critical performance indicator is the sensitivity. This latter depends on the collection efficiency, i.e. the ratio of the collected electrons to the incoming photons, and on the optical transmission of the stack. Thus it is possible to increase the sensitivity by improving this stack optical transmission without changing the photodiode. We proposed to reduce the stack height above the photodiode of a published 1.75µm pixel [17] of approximately 30%, to get a better flux concentration: the angular acceptance of the pixel is increased because the microlens base is closer to the silicon level (see Figure 11).

Our electromagnetic simulations evaluated the potential optical gain of this process at 1.25, for an aperture f#=2.8. We measured 1.21, confirming the relevance of our model.

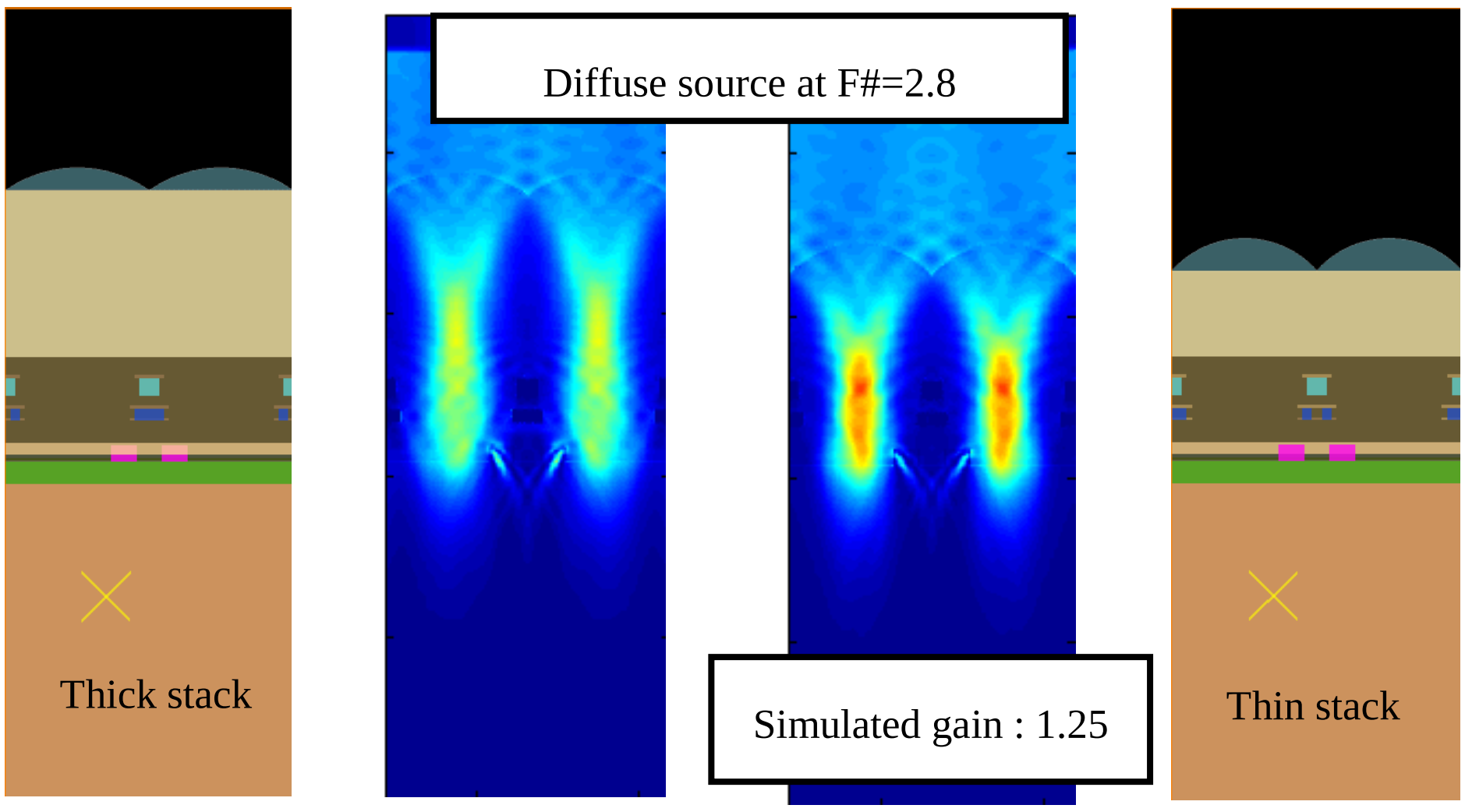

Figure 11: Light flux concentration by reducing stack height

### 3.3 Aperture effect

Regarding the miniaturization of the complete camera module, it is necessary to anticipate the pixel optical performances versus the objective-lens aperture. Indeed if one wants to keep a good image quality, it is critical to link the design of the optical package with the pixel performances. Thus we compared the optical transmission of the 1.75µm pixel, processed in the thick and thin stack presented in the section 3.2. On Figure 12 are plotted the simulated and measured normalized optical transmissions of the 2 processes as a function of the source aperture [18], the reference is the optical transmission of the thick stack at f#=2.8.

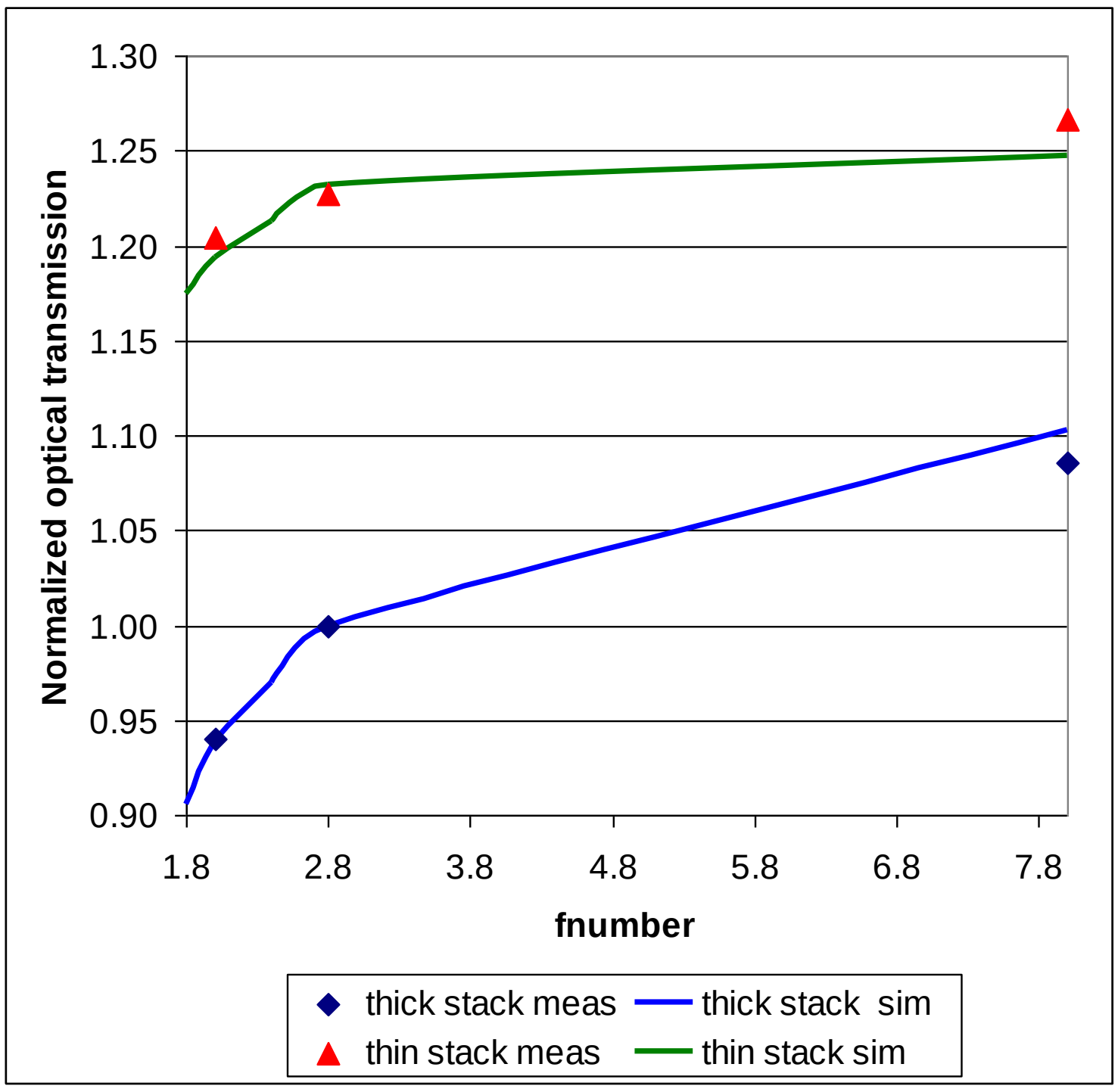

Figure 12: Influence of the aperture on the pixel optical transmission

We can see from Figure 12 that our predictions are in line with the measurements. This is an important achievement because now we are able to predict the pixel sensitivity variation with fnumber. The optical transmissions of the two stacks decrease with the fnumber whereas the sensor illuminance increases (proportional to $1/f\#^2$), which can lead to a trade-off between camera module aperture and pixel sensitivity. Moreover the thin stack optical transmission variation with fnumber remains flatter than the thick stack ones, allowing larger apertures. These predictions are essential to optimize the camera module and the sensor to follow the customers' optical performances roadmap.

## 4. CONCLUSION

Since we developed our home-made ray tracing methodology in 2002, pixels sizes have shrunk from 5.6µm to 1.75µm and below nowadays. This methodology was based on Snell-Descartes description, and was robust and efficient to predict pixels optical performances until 3µm. Unfortunately from this size, the ray-tracing models were not giving satisfying results because diffraction effects are not negligible anymore. We decided to change our model and to adopt a Maxwell-Boltzmann description based on finite-difference time domain computation. We customized this methodology to get a source model as close as possible to the product-like illumination, coupled with Bloch periodic conditions allowing us to compute these structures. Now this methodology has been validated by comparing the simulated and measured angular responses of a 1.75µm pixel. From there we proposed a robust solution to improve the sensitivity of this pixel, by reducing the stack height of 30% (thin stack), increasing the sensitivity of more than 20% in line with the simulation predictions. Finally we anticipated the aperture effects on our pixels optical performances and demonstrated that this thin stack solution was allowing larger apertures.

Now we are developing the coupling of this purely optical FDTD tool to a device simulation tool, to get a complete simulation flow from optics until photogenerated electrons and finally quantum efficiency. This approach has yet been investigated [19 and 20], but for pure monochromatic plane waves and in 2D modeling. Thus the coupling of our optics modeling tool to a device tool will allow us to predict electro-optical performances in product-like use.

## ACKNOLEDGMENTS

The author gratefully acknowledges Dr. James Pond, Chief Technology Officer of *Lumerical Solutions*, and its team, for their support in customizing the software to make this methodology applicable.

## REFERENCES

[1] L. Saurei, G. Mathieu, and B. Berge, "Design of an autofocus lens for VGA ¼-in. CCD and CMOS sensors", Proc. SPIE 5249, 288-296 (2004)

[2] H. Ren and S. T. Wu, "Variable-focus liquid lens**"**, Optics Express, Vol. 15, Issue 10, pp. 5931-5936, 2007

[3] H. Ren, D. W. Fox, B. Wu, and S. T. Wu, "Liquid crystal lens with large focal length tunability and low operating voltage", Optics Express, Vol. 15, Issue 18, pp. 11328-11335, August 2007.

[4] S.-.Y. Lee, H.-W. Tung, W.-C. Chen and W. Fang, "Thermal actuated tunable lens", IEEE Photonics Technology Letters, vol.18, NO. 21, November 2006.

[5] E. Fossum, "Active Pixel Sensors: Are CCD's Dinosaurs?", Charge-Coupled Devices and Solid State Optical Sensors III, Morley M. Blouke; Ed., Proc. SPIE Vol. 1900, pp 2-14, July 1993

[6] J. Janesick, "Lux transfer: Complementary metal oxide semiconductors versus charge-coupled devices", Optical Engineering, **41** (06), pp 1203-1215., June 2002

[7] E. Fossum, "CMOS image sensors: Electronic Camera-On-A-Chip", *IEEE* Transactions on electron devices, **44** (10), pp 1689-1698, Oct. 1997.

[8] B. E. Bayer, Color Imaging Array, In United States Patents, number 3971065, Eastman Kodak Company, Rochester, N.Y., July 1976

[9] http://www.zemax.com

[10] J. Vaillant, F. Hirigoyen, "Optical simulation for CMOS imager microlens optimization", Proc. of the SPIE, Volume 5459, pp. 200-210, September 2004.

[11] A. Taflove and S. C. Hagness, "Computational Electrodynamics : the finite-difference time-domain method, 2nd Edition", H. E. Schrank, Series Editor (Artech House, Boston, Ma, 2000).

[12] K. S. Yee, "Numerical solution of initial boundary value problems involving Mawwell's equations in isotropic media", IEEE Transactions on Antennas and Propagation, Vol. AP-14, number 3, pp. 302-307, 1966.

[13] A. Taflove and M. E. Brodwin (1975). "Numerical solution of steady-state electromagnetic scattering problems using the time-dependent Maxwell's equations". *Microwave* Theory and Technique,, IEEE Transactions on 23: 623–630

[14] http://www.lumerical.com

[15] A.Crocherie, J. Vaillant, F. Hirigoyen, “3D Broadband FDTD optical simulations of CMOS image sensor” submitted in IEEE Transactions on Electron Device.

[16] J. Vaillant, A. Crocherie, F. Hirigoyen, A. Cadien, J. Pond, “Uniform Illumination and rigorous electromagnetic simulations applied to CMOS Image Sensors”, *Optics Express*, Vol. 15, Issue 9, pp. 5494-5503, April 2007

[17] M. Cohen *et al.*, “Fully Optimized Cu based process with dedicated cavity etch for 1.75µm and 1.45µm pixel pitch CMOS Image Sensors”, International Electron Devices Meeting IEDM’06.

[18] J. Vaillant *et al.*, “Versatile method for optical performances characterization of off-axis CMOS pixels with microlens radial shift”, Proc. SPIE, 6817 (2008).

[19] C. H. Koo, *et. al,* “Improvement of Crosstalk on 5M CMOS Image Sensor with 1.7x1.7µm² pixels”, Proc. SPIE, 6471 (2007).

[20] S. H. Cho *et al.*,”Optoelectronic Investigation for High Performance 1.4µm Pixel CMOS Image Sensors”, Proc. Int. Image Sensor Workshop, pp.13–15, Maine, June 2007.